\documentclass[pre,showpacs,amsmath,amssymb]{revtex4-1}

\usepackage{graphicx}
\usepackage{amsmath}

\newcommand{\Dlt}{{D_{\rm L,\rm T}}}
\newcommand{\Dl}{{D_{\rm L}}}
\newcommand{\Dt}{{D_{\rm T}}}

\newcommand{\etam}{\eta_{\rm m}}
\newcommand{\etaeff}{\eta^{\rm eff}}
\newcommand{\etaf}{\eta_{\rm f}}
\newcommand{\etameff}{\eta_{\rm m}^{\rm eff}}
\newcommand{\ie}{{i.e., }}
\newcommand{\kT}{{k_{\rm B}T}}

\newcommand{\pd}{{\partial}}
\newcommand{\rhom}{\rho_{\rm m}}
\newcommand{\tenG}{{\bf G}}
\newcommand{\tG}{\tilde{G}}

\newcommand{\vecF}{{\bf F}}
\newcommand{\veck}{{\bf k}}
\newcommand{\vecq}{{\bf q}}
\newcommand{\vecr}{{\bf r}}
\newcommand{\vecv}{{\bf v}}
\newcommand{\xhat}{\hat{x}}

\newcommand{\tenS}{{\bf S}}

\begin{document}

\title{Correlated dynamics of inclusions in a supported membrane}

\author{Naomi Oppenheimer}
\author{Haim Diamant}
\email{hdiamant@tau.ac.il}
\homepage{http://www.tau.ac.il/~hdiamant}
\affiliation{Beverly and Raymond Sackler School of Chemistry,
Tel Aviv University, Tel Aviv 69978, Israel}

\date{August 1, 2010}

\begin{abstract}
  
  The hydrodynamic theory of heterogeneous fluid membranes is extended
  to the case of a membrane adjacent to a solid substrate. We derive
  the coupling diffusion coefficients of pairs of membrane inclusions
  in the limit of large separation compared to the inclusion size.
  Two-dimensional compressive stresses in the membrane make the
  coupling coefficients decay asymptotically as $1/r^2$ with
  interparticle distance $r$. For the common case, where the distance
  to the substrate is of sub-micron scale, we present expressions for
  the coupling between distant disklike inclusions, which are valid
  for arbitrary inclusion size. We calculate the effect of inclusions
  on the response of the membrane and the associated corrections to
  the coupling diffusion coefficients to leading order in the
  concentration of inclusions. While at short distances the response
  is modified as if the membrane were a two-dimensional suspension,
  the large-distance response is not renormalized by the inclusions.

\end{abstract}


\maketitle
\section{introduction}
\label{sec_intro}

Biological membranes are fluid bilayers made primarily of lipid
molecules \cite{bio}. From a hydrodynamic point of view such a bilayer
is a quasi-two-dimensional (quasi-2D) viscous liquid, whose molecules
are constrained to flow along the 2D membrane surface while exchanging
momentum not only among themselves but also with the surrounding
three-dimensional (3D) solvent. Biomembranes contain also a high
concentration of embedded inclusions --- integral proteins and
possibly also nanometer-scale domains --- which perform key biological
functions and are typically much larger than the lipids \cite{bio}.
Thus, from the same coarse-grained perspective, a biomembrane can be
viewed as a quasi-2D suspension \cite{jpsj09}. We have recently used
this perspective to investigate the correlated motion of proteins in a
membrane freely floating in an unbounded liquid, and how the
inclusions affect the response of such a membrane to stresses
\cite{bpj09}. In many practical circumstances, however, the membrane
is not free but attached to a solid substrate, such as the elastic
scaffold of the cytoskeleton or an external surface to which a cell
adheres. Additionally, in various experiments membranes are more
easily studied when supported by a solid substrate. The aim of the
current work, therefore, is to explore how the results of Ref.\
\cite{bpj09} are modified by the presence of such a nearby immobile
surface.

Two major approaches to the hydrodynamics of free membranes have been
presented. The first, by Saffman and Delbr\"uck (SD) \cite{SD}, models
the membrane as a viscous liquid slab of width $w$ and viscosity
$\etam/w$, having no-slip contacts at its bounding surfaces with two
semi-infinite fluids of viscosity $\etaf$. The second, by Levine et
al.\ \cite{Levine}, considers the membrane as a vanishingly thin
viscoelastic film embedded in an infinite viscous fluid.  The dynamics
of membranes embedded in a 3D fluid have been studied also using
computer simulations \cite{Brown,ShigeEPL,Brown2}. The common key feature of
these theories is the fact that the membrane does not conserve
momentum in 2D, while the total momentum is conserved in 3D.
Consequently, a length scale $\kappa^{-1}$ emerges, characterizing the
crossover from a 2D-like membrane response, where stresses dominantly
propagate through the membrane, to a 3D-like response, where the outer
fluid governs the dynamics. This length is determined by the ratio
between the 2D viscosity of the membrane and the 3D viscosity of the
surrounding fluid \cite{SD}.  For lipid bilayers $\kappa^{-1}$ is
typically two to three orders of magnitude larger than the membrane
thickness $w$, \ie of micron scale. The work of Ref.\ \cite{bpj09} is
an extension of the SD theory to cases with more than one inclusion,
where there are three lengths to consider: the lateral size (radius)
$a$ of the inclusion, the SD length $\kappa^{-1}$, and the distance
$r$ between the inclusions. The analysis was restricted to the limit
$a\ll\min(\kappa^{-1},r)$, where complications related to specific
details of the inclusions \cite{Naji} are immaterial. We employ the
same assumption in most of the current work as well.

The introduction of an immobile surface breaks the translational
symmetry in the directions parallel to the membrane and, hence,
qualitatively changes the hydrodynamics of the system as its total
momentum is no longer conserved \cite{jpsj09}. Models appropriate for
such a scenario can be divided into two groups. The first
\cite{Mazenko,ES,Suzuki,Komura,Nelson} adopts a phenomenological
approach, describing the membrane as a 2D Brinkman fluid
\cite{Brinkman}, \ie introducing a term in the 2D hydrodynamic
equation for the membrane, which leaks momentum at a certain fixed
rate $\alpha^2\nu_{\rm m}$ ($\nu_{\rm m}=\etam/\rhom$ being the 2D
kinematic viscosity of the membrane and $\rhom$ its 2D mass density).
This sets a phenomenological screening length, $\alpha^{-1}$, beyond
which momentum is lost to the substrate. In the second approach
\cite{Lubensky,SA,Fischer,ShigeEPJE} the additional length scale is
explicitly determined by the thickness $h$ of a fluid layer separating
the membrane from the solid substrate. When fluid exists only between
the membrane and the substrate (as in the case of a supported lipid
monolayer) \cite{Lubensky,SA,Fischer}, or when the membrane lies at
the midplane between two substrates \cite{ShigeEPL,ShigeEPJE}, the
large-distance effects converge to those of the phenomenological,
Brinkman-like approach in the limit $h\ll\kappa^{-1}$ (\ie for $h$
much smaller than a micron). The momentum screening length thus
obtained is the geometrical mean of the other two lengths,
$\alpha^{-1}\sim(\kappa^{-1}h)^{1/2}$ \cite{SA}.

The current work extends the analysis of Refs.\
\cite{Lubensky,SA,Fischer} to the realistic scenario of a supported
membrane with fluid on both sides, containing more than one
inclusion. In Sec.\ \ref{sec_model} we define the model and present
the results for the hydrodynamics of a supported membrane, which will
be useful for our analysis. Because of the three length scales in the
model --- $\kappa^{-1}$, $h$, and the interparticle distance $r$
(assuming that $a$ is much smaller than all three) --- there are
several asymptotic regimes to be considered, which are defined and
discussed in Sec.\ \ref{sec_model}.  These results are used in Sec.\
\ref{sec_diffusion} to calculate the coupling diffusion coefficients
of pairs of inclusions, which should be directly measurable in
particle-tracking experiments.  We proceed in Secs.\
\ref{sec_effective} and \ref{sec_corrected} to examine the effect of a
finite concentration of inclusions on the response of the supported
membrane to stresses (\ie the effective viscosity of the membrane) and
the resulting corrections to the coupling diffusion coefficients. In
the practically useful limit of $h\ll\kappa^{-1}$ we have been able to
derive the large-separation coupling diffusion coefficients for
inclusions of arbitrary size; these results are presented in Sec.\
\ref{sec_large}.  The conclusions are summarized in Sec.\
\ref{sec_conclusion}.

\section{Model}
\label{sec_model}

\begin{figure}[tbh]
\vspace{0.7cm} 
\centerline{\resizebox{0.6\textwidth}{!}{\includegraphics{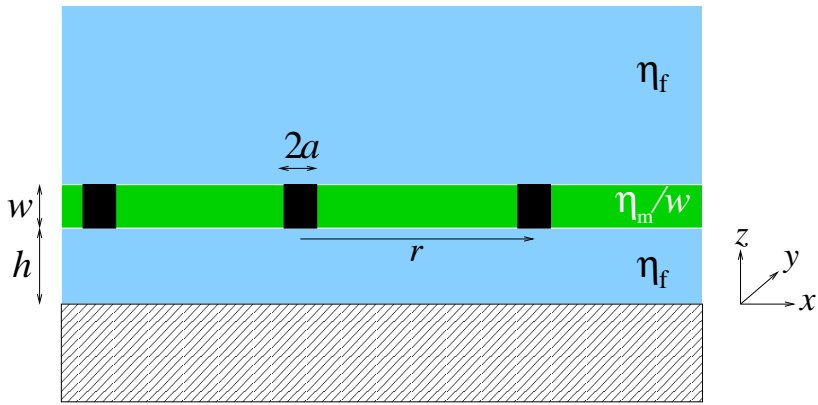}}}
\caption{(Color online). 
Schematic view of the model system and its parametrization.}
\label{fig_system}
\end{figure}

Our model system is similar to that of Refs.\
\cite{Lubensky,SA,Fischer} and is schematically depicted in Fig.\
\ref{fig_system}. A flat slab of viscous liquid (width $w$ and
viscosity $\etam/w$) lies on the $xy$ plane a distance $h$ away from a
flat rigid surface. The space between the slab and the surface is
filled with another fluid (viscosity $\etaf$).  Unlike the system
treated in Refs.\ \cite{Lubensky,SA,Fischer}, the space on the other
side of the slab is occupied by a semi-infinite fluid of viscosity
$\etaf$ as well. All fluids are assumed incompressible. The SD length
is defined as
\begin{equation}
  \kappa^{-1} = \frac{\etam}{2\etaf}.
\label{SD}
\end{equation}
In the slab are rigid inclusions of radius $a$. Although they are
depicted as cylinders, their exact shape does not affect most of our
analysis; in points where it does, this will be explicitly mentioned.
We do not consider effects related to curvature and thermal
fluctuations of the membrane \cite{Brown,SunilKumar,GovZilman,Seifert}.

Our goal is to characterize the response of the inclusion-decorated
membrane to stresses and relate it to the coupled motions of two
inclusions. A common way to characterize the membrane response as a
fluid medium is through the velocity Green's function, $\tenG(\vecr)$.
This tensor gives the flow velocity $\vecv(\vecr)$ of the membrane at
the 2D position $\vecr$ due to a point force $\vecF$ exerted on the
membrane at the origin in the $xy$ plane, according to
$v_i(\vecr)=G_{ij}(\vecr)F_j$ (with $i,j=x,y$ and summation over the
repeated index $j$). When the separation between two inclusions is
much larger than their sizes ($r\gg a$), the pair mobility and pair
diffusion coefficients associated with their coupled motions can be
directly obtained from the velocity Green's function, as explained in
detail in Sec.\ \ref{sec_diffusion}.

In unbounded 3D liquids $\tenG$ is the Oseen tensor \cite{Faxen}. We
need the analogue of the Oseen tensor for the model system of Fig.\ 
\ref{fig_system}. The velocity Green's function for a similar model of
a supported monolayer, where the upper fluid is absent, was derived in
Ref.\ \cite{Lubensky}. Its generalization to the case of two different
fluids and two confining surfaces has been performed in Ref.\ 
\cite{ShigeEPJE}. We use this result while specializing to two
identical outer fluids, the upper one being semi-infinite. In Fourier
space [$\tilde{\tenG}(\vecq)=\int d^2r
e^{-i\vecq\cdot\vecr}\tenG(\vecr)$] the Green's function is given by
\cite{ShigeEPJE}
\begin{equation}
  \tG_{ij}(\vecq) = \frac{1}{\etaf q [\coth(qh) + 2q/\kappa + 1]}
  \left( \delta_{ij} - \frac{q_iq_j}{q^2} \right).
\label{Gq}
\end{equation}

In the limit of an infinitely distant surface Eq.\ (\ref{Gq})
coincides with the Green's function for a free membrane
\cite{SD,Levine,bpj09},
\begin{equation}
  \tG_{ij} \overset{h\rightarrow\infty}{\longrightarrow}
  \tG^{\rm f}_{ij}(\vecq) = \frac{1}{\etam q (q + \kappa)}
  \left( \delta_{ij} - \frac{q_iq_j}{q^2} \right).
\label{Gqf}
\end{equation}
In the other limit, of a vanishingly small $h$, it becomes
\begin{equation}
 \tG_{ij} \overset{h\rightarrow 0}{\longrightarrow}
  \frac{1}{\etam (q^2 + \kappa q/2 + \alpha^2)}
  \left( \delta_{ij} - \frac{q_iq_j}{q^2} \right),
\label{Gqa1}
\end{equation}
with $\alpha=[\kappa/(2h)]^{1/2}$.  The function in Eq.\ (\ref{Gqa1})
has two poles at
$q_\pm=\kappa[-1\pm\sqrt{1-8/(\kappa h)}]/4$,
which in the limit $h\rightarrow 0$ turn into $q_\pm=\pm
i\alpha$. Hence, in this limit $\tilde{\tenG}$ coincides with the 2D
Brinkman-like Green's function for an adsorbed membrane
\cite{Mazenko,ES,Suzuki,Komura,Nelson},
\begin{equation}
   \tG_{ij} \overset{h\rightarrow 0}{\longrightarrow} \tG^{\rm a}_{ij}(\vecq)
  = \frac{1}{\etam (q^2 + \alpha^2)}
  \left( \delta_{ij} - \frac{q_iq_j}{q^2} \right).
\label{Gqa}
\end{equation}
Thus, in the strongly adsorbed limit the presence of the upper fluid
merely adds a prefactor to the momentum screening length compared to
the monolayer case studied earlier \cite{SA}
[$\alpha^{-1}=(2\kappa^{-1}h)^{1/2}$ instead of
$(\kappa^{-1}h)^{1/2}$].

The Green's function of Eq.\ (\ref{Gq}) serves as the starting point for
the entire analysis to follow. It is beneficial, therefore, to begin
by exploring the different asymptotic regimes that $\tenG(\vecr)$
defines. Since it depends on three lengths --- $\kappa^{-1}$, $h$, and
$r$ --- there are quite a few such regimes. Let us first rewrite it as
a function of two length ratios,
\begin{eqnarray}
  G_{ij}(\vecr) &=& \frac{1}{\etaf h} g_{ij}(\kappa^{-1}/h,\vecr/h) \nonumber\\
  g_{ij} &=& \frac{1}{(2\pi)^2} \int d^2k e^{i\veck\cdot(\vecr/h)} 
   \frac{1}{k[\coth k +(2\kappa^{-1}/h)k + 1]} \left( \delta_{ij} - \frac{k_ik_j}{k^2} \right),
\label{gij}
\end{eqnarray}
thus making the asymptotics more transparent.

\subsection{Adsorbed regime: $h\ll\kappa^{-1}$} 
\label{sec_model_a}

In this commonly encountered regime, where the distance to the
substrate is much smaller than the SD length, the integral in Eq.\ 
(\ref{gij}) is governed for all values of $r$ by small $k$, whereby
$\coth k+(2\kappa^{-1}/h)k \simeq 1/k + (2\kappa^{-1}/h)k$.  Due to
the argument presented below Eq.\ (\ref{Gqa1}), the Green's function
coincides with $\tilde{\tenG}^{\rm a}$ of Eq.\ (\ref{Gqa}).  Inverting
it to real space, one gets
\begin{eqnarray}
  G^{\rm a}_{ij}(\vecr) &=& \frac{1}{2\pi\etam} \left[
    \frac{1}{2} [ K_0(\alpha r) + K_2(\alpha r) ] \delta_{ij}
    - K_2(\alpha r) \frac{r_i r_j}{r^2}
    \right] \nonumber \\
  &-&\frac{1}{2\pi\etam\alpha^2 r^2} \left( 
  \delta_{ij} - \frac{2r_ir_j}{r^2} \right),
\label{Gra}
\end{eqnarray}
where $K_n$ are modified Bessel functions of the second kind. The
first term in Eq.\ (\ref{Gra}) is short-ranged, decaying exponentially
with $\alpha r$. It arises from shear stresses in the membrane, which
get screened for $r>\alpha^{-1}$. The second term is long-ranged,
decaying only as $1/r^2$ and originating from compressive stresses in
the membrane. The effect of such stresses on the steady flow is that
of an effective 2D mass dipole \cite{jpsj09}, whose magnitude is
proportional to $(\etam\alpha^2)^{-1}\sim h/\etaf$, independent of
membrane viscosity.

Thus, the adsorbed regime is subdivided into two regions reflecting
different physics. In the {\it adsorbed near} region, $r\ll
\alpha^{-1}=(2\kappa^{-1}h)^{1/2}\ll\kappa^{-1}$, the response is
governed by the yet-unscreened 2D shear stresses in the membrane.
Expanding Eq.\ (\ref{Gra}) in small $\alpha r$, we have
\begin{equation}
  G_{ij}(\vecr) \simeq G^{\rm an}_{ij}(\vecr) = 
  \frac{1}{4\pi\etam} \left\{ -\left[ \ln(\alpha r/2) + \gamma + 1/2 \right]
  \delta_{ij} + \frac{r_ir_j}{r^2} \right\} + O[(\alpha r)^2/\etam],
\label{Gan}
\end{equation}
where $\gamma$ is Euler's constant. This result coincides with the
Oseen tensor of a momentum-conserving 2D liquid, exhibiting the well
known logarithmic behavior of this problem, with a cutoff length of
$\alpha^{-1}$.
In the {\it adsorbed far} region,
$r\gg\alpha^{-1}=(2\kappa^{-1}h)^{1/2}\gg h$, the response is due to
long-ranged compressive stresses, yielding
\begin{equation}
  G_{ij}(\vecr) \simeq G^{\rm af}_{ij}(\vecr) =  -\frac{h}{2\pi\etaf r^2} \left( 
  \delta_{ij} - \frac{2r_ir_j}{r^2} \right) + O[(\alpha r)^{-1/2}e^{-\alpha r}/\etam].
\label{Gaf}
\end{equation}
Note that, since in the adsorbed regime $h<\alpha^{-1}<\kappa^{-1}$,
the adsorbed near region includes distances $r$ both smaller and
larger than $h$, and the adsorbed far one includes distances
both smaller and larger than $\kappa^{-1}$.

\subsection{Hovering regime: $\kappa^{-1}\ll h$} 
\label{sec_model_h}

In this regime, where the thickness $h$ of the fluid layer between the
membrane and the substrate is the larger length scale, the asymptotes
of Eq.\ (\ref{gij}) depend on the value of $r/h$. For $r\ll h$ the
integral in Eq.\ (\ref{gij}) is governed by large $k$, whereby $\coth
k+(2\kappa^{-1}/h)k \simeq 1 + (2\kappa^{-1}/h)k$.  The Green's function
then coincides with that of a free membrane, $\tilde{\tenG}^{\rm f}$
of Eq.\ (\ref{Gqf}). This is the case studied in Ref.\
\cite{bpj09}. Inverting $\tilde{\tenG}^{\rm f}$ to real space
\cite{Levine,bpj09} yields
\begin{eqnarray}
  G^{\rm f}_{ij}(\vecr) &=& \frac{1}{4\etam} \left\{
    \left[ H_0(\kappa r)-\frac{H_1(\kappa r)}{\kappa r}
    -\frac{1}{2}\left(Y_0(\kappa r)-Y_2(\kappa r)\right) +
    \frac{2}{\pi(\kappa r)^2} \right] \delta_{ij} \right. \nonumber\\
    &-& \left. \left[ H_0(\kappa r)-\frac{2H_1(\kappa r)}{\kappa r} + Y_2(\kappa r)
  + \frac{4}{\pi(\kappa r)^2} \right] \frac{r_ir_j}{r^2} \right\},  
\label{Grf}
\end{eqnarray}
where $Y_n$ are Bessel functions of the second kind, and $H_n$ Struve
functions. 

The free behavior is subdivided into two regions having different
physics. For $r\ll\kappa^{-1}\ll h$ we have the {\it free near}
region, where Eq.\ (\ref{Grf}) becomes
\begin{equation}
  G_{ij}(\vecr) \simeq G^{\rm fn}_{ij}(\vecr) =
  \frac{1}{4\pi\etam} \left\{ -\left[ \ln(\kappa r/2) + \gamma + 1/2 \right]
  \delta_{ij} + \frac{r_ir_j}{r^2} \right\}
  + O(\kappa r/\etam).
\label{Gfn}
\end{equation}
As in the adsorbed near region [Eq.\ (\ref{Gan})] the response in the
free near region is governed by 2D shear stresses in the membrane. The
difference is in the cutoff length, which in this case is
$\kappa^{-1}$ rather than $\alpha^{-1}$.  In the {\it free far}
region, $\kappa^{-1}\ll r\ll h$, Eq.\ (\ref{Grf}) yields
\begin{equation}
  G_{ij}(\vecr) \simeq G^{\rm ff}_{ij}(\vecr) = \frac{1}{4\pi\etaf}\frac{r_ir_j}{r^3}
 + O[(\kappa r)^{-2}/\etam]. 
\label{Gff}
\end{equation}
Both the $1/r$ decay and the dependence on $\etaf$ rather than $\etam$
indicate that the response in this region is due to shear stresses in
the 3D fluid on the two sides of the membrane. 

In the last asymptotic region, $\kappa^{-1}\ll h\ll r$, although $h$
is the larger characteristic length, the distance $r$ is sufficiently
large to make the fact that the membrane is supported rather than free
come into play.  In this {\it supported} region the integral in Eq.\
(\ref{gij}) is governed by small $k$, yet, unlike the adsorbed regime,
$\coth k + (2\kappa^{-1}/h)k \simeq 1/k+k/3$, independent of
$\kappa$. The resulting Green's function is
\begin{equation}
  \tG_{ij}(\vecq) \simeq \tG^{\rm s}_{ij}(\vecq) = \frac{1}
   {\etaf h (q^2/3 +q/h+ h^{-2})}
  \left( \delta_{ij} - \frac{q_iq_j}{q^2} \right),
\label{Gqs}
\end{equation}
which has two poles, both depending on $h$ alone --- \ie $h$ is
the sole momentum screening length in this region. Inverting Eq.\ 
(\ref{Gqs}) to real space and taking the limit $r\gg h$ yields
\begin{equation}
  G_{ij}(\vecr) \simeq G^{\rm s}_{ij}(\vecr) = -\frac{h}{2\pi\etaf r^2} \left( 
  \delta_{ij} - \frac{2r_ir_j}{r^2} \right) + O[h^2/(\etaf r^3)].
\label{Gsup}
\end{equation}
Thus, despite differences in the details ($\alpha^{-1}$ vs.\ $h$ as
the length scale for momentum loss), the asymptotic responses in the
adsorbed far and supported regions [Eqs.\ (\ref{Gaf}) and
(\ref{Gsup}), respectively] turn out to be identical, both arising
from 2D compressive stresses due to an effective mass dipole of
magnitude $\sim h/\etaf$. Note, however, that the correction to the
leading mass-dipole term in Eq.\ (\ref{Gaf}) is exponentially small,
whereas in the supported region the correction is algebraic. This
reflects the fact that in the adsorbed region the upper fluid is
insignificant, whereas in the supported region it does play a role,
albeit not a dominant one.

The various spatial regimes are summarized in Table \ref{tab_regime}.

\begin{table}[tbh]
\begin{ruledtabular}
\begin{tabular}{llccl}
regime & sub-region & definition & spatial dependence & mechanism \\
\hline
adsorbed & & $h\ll\kappa^{-1}$ & & \\
& near & $r\ll\alpha^{-1}\equiv(2\kappa^{-1}h)^{1/2}$ & $\ln(\alpha r)$ & 2D shear \\
& far & $r\gg\alpha^{-1}$ & $r^{-2}$ & 2D compression \\
\hline
hovering & & $\kappa^{-1}\ll h$ & \\
& free near & $r\ll\kappa^{-1}$ & $\ln(\kappa r)$ & 2D shear \\
& free far & $\kappa^{-1}\ll r\ll h$ & $r^{-1}$ & 3D shear \\
& supported & $r\gg h$ & $r^{-2}$ & 2D compression
\end{tabular}
\end{ruledtabular}
\caption{Summary of asymptotic spatial regimes and their corresponding notation.}
\label{tab_regime}
\end{table}

\section{Correlated diffusion}
\label{sec_diffusion}

The Green's function of Eq.\ (\ref{gij}) gives the membrane flow
velocity at position $\vecr$ in response to a unit force exerted on it
at the origin. In the limit $r\gg a$, addressed in this article, the
same Green's function also gives the coupling mobility tensor,
$B_{12,ij}(\vecr)$ --- \ie the velocity of one particle due to a unit
force acting on another, where the positions of the two particles are
separated by the vector $\vecr$, $v_{1,i}=B_{12,ij}(\vecr)F_{2,j}$
(with summation over the repeated index $j$). From the mobility tensor
the coupling diffusion tensor, $D_{12,ij}(\vecr)$, readily follows via
the Einstein relation, $D_{12,ij}=\kT B_{12,ij}$, $\kT$ being the
thermal energy.  The $x$ axis can be defined, without loss of
generality, along the line connecting the pair, $\vecr=r\xhat$.  This
choice leads, by symmetry, to $D_{12,xy}=0$.  The coupled diffusion of
the two particles is then fully characterized by two coefficients: a
longitudinal coupling diffusion coefficient,
$\Dl(r)=D_{12,xx}(r\xhat)$, and a transverse one,
$\Dt(r)=D_{12,yy}(r\xhat)$. Thus, in summary, we have
\begin{equation}
  \Dl(r\gg a) = \kT G_{xx}(r\xhat),\ \ \ 
  \Dt(r\gg a) = \kT G_{yy}(r\xhat).
\label{DltG}
\end{equation}
The first coefficient is associated with the coupled Brownian motion
of the pair along their connecting line, while the second --- to the
coupled motion perpendicular to that line,
\begin{equation}
  \langle \Delta x_1\Delta x_2\rangle = 2\Dl(r)t,\ \ \ 
  \langle \Delta y_1\Delta y_2\rangle = 2\Dt(r)t,
\label{fluct}
\end{equation}
where $\Delta x_\beta,\Delta y_\beta$ ($\beta=1,2$) are the
displacements of particle $\beta$ during time $t$.  We shall now give
the expressions for these coupling diffusion coefficients in the
various asymptotic regimes (cf.\ Table \ref{tab_regime}).

\subsection{Adsorbed regime: $h\ll\kappa^{-1}$} 
\label{sec_diffusion_a}

In the adsorbed regime we get from Eqs.\ (\ref{Gra}) and (\ref{DltG})
the following coupling diffusion coefficients:
\begin{eqnarray}
  h\ll\kappa^{-1}:\ \ && \Dl(r) \simeq \frac{\kT}{2\pi\etam} \left[
  -\frac{K_1(\alpha r)}{\alpha r} + \frac{1}{\alpha^2r^2} \right] 
 \nonumber\\
  && \Dt(r) \simeq \frac{\kT}{2\pi\etam} \left[
  K_0(\alpha r) + \frac{K_1(\alpha r)}{\alpha r} - \frac{1}{\alpha^2r^2} \right].
\label{Dlta}
\end{eqnarray}
This regime is subdivided into near and far regions. In the adsorbed near region
Eqs.\ (\ref{Gan}) and (\ref{DltG}) yield
\begin{equation}
 h\ll\kappa^{-1}:\ \ \Dlt(r\ll\alpha^{-1}) \simeq 
  \frac{\kT}{4\pi\etam} \left[ -\ln(\alpha r/2) - \gamma \pm 1/2 \right],
\label{Dlt_an}
\end{equation}
whereas in the adsorbed far region we obtain from Eqs.\ (\ref{Gaf})
and (\ref{DltG})
\begin{equation}
  h\ll\kappa^{-1}:\ \ \Dlt(r\gg\alpha^{-1}) \simeq 
  \pm\frac{\kT h}{2\pi\etaf r^2}.
\label{Dlt_af}
\end{equation}
In Eqs.\ (\ref{Dlt_an}) and (\ref{Dlt_af}) the upper (lower) sign
corresponds to the longitudinal (transverse) coefficient.  Notice how
at large distances, $r\gg\alpha^{-1}$, the coupling diffusion coefficients
both decay as $1/r^2$, suggesting the dominance of the aforementioned
mass dipole. Momentum at such distances is lost to the solid substrate
and the coupling is mediated by 2D compressive stresses in the
membrane.

\begin{figure}[tbh]
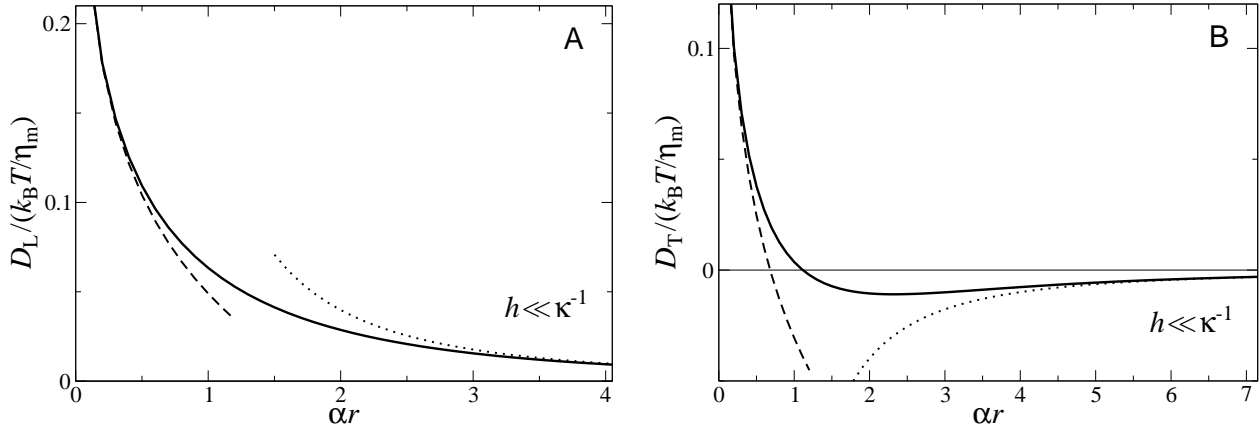

\vspace{0.7cm} 
\centerline{\resizebox{0.45\textwidth}{!}{\includegraphics{fig2a.eps}}
\hspace{0.3cm}
\resizebox{0.45\textwidth}{!}{\includegraphics{fig2b.eps}}}
\caption{Longitudinal (A) and transverse (B) coupling diffusion coefficients 
  as a function of interparticle distance for the adsorbed regime
  ($h\ll\kappa^{-1}$). Diffusion coefficients are scaled by
  $\kT/\etam$ and the distance by the momentum decay length
  $\alpha^{-1}$. The full behavior [Eq.\ (\ref{Dlta}), solid] is shown
  together with those in the two asymptotic regions: adsorbed near
  region ($\Dlt\sim\ln r$, dashed); adsorbed far region ($\Dlt\sim\pm
  1/r^2$, dotted).}
\label{fig_Da}
\end{figure}

The dependencies of the coupling diffusion coefficients on the
separation between the inclusions for the adsorbed regime are shown in
Fig.\ \ref{fig_Da} along with their various asymptotes.

\subsection{Hovering regime: $\kappa^{-1}\ll h$} 
\label{sec_diffusion_h}

The hovering regime is divided into the free ($r\ll h$) and supported
($r\gg h$) behaviors. In the free limit we get from Eqs.\ (\ref{Grf})
and (\ref{DltG})
\begin{eqnarray}
  \kappa^{-1}\ll h:\ \  
   && \Dl(r\ll h) \simeq \frac{\kT}{4\etam\kappa r} \left\{ H_1(\kappa r) - Y_1(\kappa r)
    -\frac{2}{\pi\kappa r} \right\}
 \nonumber\\
  && \Dt(r\ll h) \simeq \frac{\kT}{4\etam} \left\{ H_0(\kappa r) - 
  \frac{H_1(\kappa r)}{\kappa r} - \frac{1}{2} \left[ Y_0(\kappa r)-Y_2(\kappa r) \right] 
  + \frac{2} {\pi\kappa^2r^2} \right\}.
\label{Dlt_f}
\end{eqnarray}
This behavior is further subdivided into near and far regions.  In the
free near region Eqs.\ (\ref{Gfn}) and (\ref{DltG}) yield the coupling
diffusion coefficients as
\begin{equation}
  \kappa^{-1}\ll h:\ \
  \Dlt (r\ll\kappa^{-1}) \simeq \frac{\kT}{4\pi\etam} \left[ -\ln(\kappa r/2) 
  - \gamma \pm 1/2 \right],
\label{Dlt_fn}
\end{equation}
where, again, the upper (lower) sign corresponds to the longitudinal
(transverse) coefficient.  Note the similarity between this result and
the one in the adsorbed near region [Eq.\ (\ref{Dlt_an})].  In both
cases the coupling is governed by the behavior of the membrane as a 2D
fluid, the only difference being the cutoff of the logarithmic term.
In the free far region Eqs.\ (\ref{Gff}) and (\ref{DltG}) lead to
\begin{eqnarray}
  \kappa^{-1}\ll h:\ \ 
  && \Dl(\kappa^{-1}\ll r\ll h) \simeq \frac{\kT} {2\pi\etam\kappa r} 
  = \frac{\kT}{4\pi\etaf r} 
 \nonumber\\
  &&\Dt(\kappa^{-1}\ll r\ll h) \simeq \frac{\kT}{2\pi\etam\kappa^2r^2} 
  = \frac{\kT\etam}{8\pi\etaf^2 r^2}.
 \label{Dlt_ff}
\end{eqnarray}
The coupling in this region is mediated by the outer 3D fluid, as
reflected by the dependence of $\Dl$ on $\etaf$ and its spatial decay
as $1/r$. The transverse coefficient decays faster (as $1/r^2$), since
it arises from an effective force dipole proportional to
$\kappa^{-1}\sim\etam/\etaf$ \cite{bpj09}. This also leads to an
unusual {\it increasing} dependence of $\Dt$ on membrane viscosity.  All of
the equations in the free limit, Eqs.\ (\ref{Dlt_f})--(\ref{Dlt_ff}),
coincide with those for a free membrane as derived in Ref.\ 
\cite{bpj09}.

In the last region, the supported region, Eqs.\ (\ref{Gsup}) and
(\ref{DltG}) give the coupling diffusion coefficients
\begin{equation}
  \kappa^{-1}\ll h:\ \
  \Dlt(r\gg h) \simeq \pm\frac{\kT h}{2\pi\etaf r^2},
\label{Dlt_hs}
\end{equation}
which are identical to those in the adsorbed far region [Eq.\ 
(\ref{Dlt_af})], as they arise in both cases from the same physical
mechanism (2D compressive stresses in the membrane).

\begin{figure}[tbh]
\vspace{0.7cm} 
\centerline{\resizebox{0.45\textwidth}{!}{\includegraphics{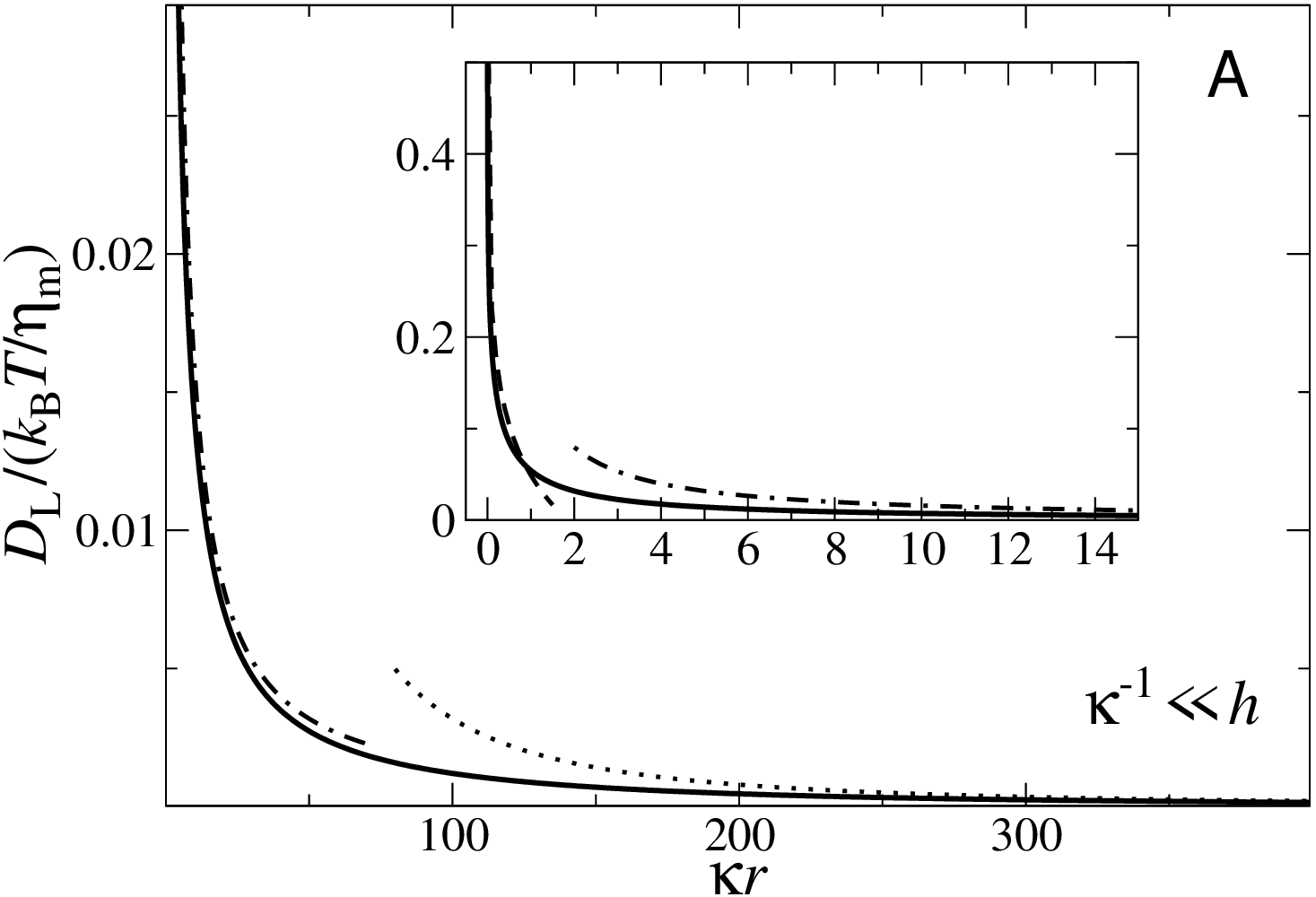}}
\hspace{0.3cm}
\resizebox{0.46\textwidth}{!}{\includegraphics{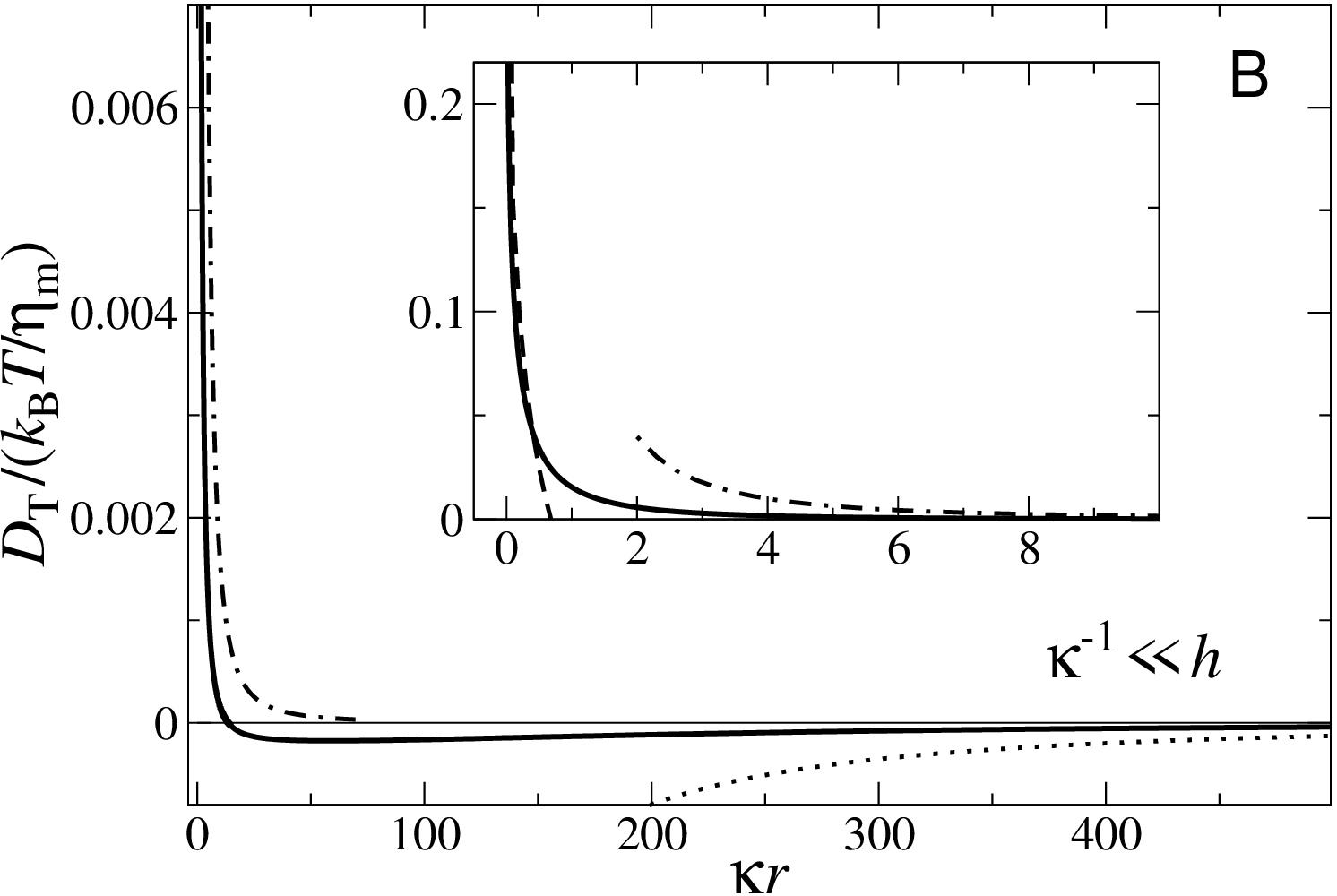}}}
\caption{Longitudinal (A) and transverse (B) coupling diffusion coefficients 
  as a function of interparticle distance for the hovering regime
  ($\kappa^{-1}\ll h$). Insets focus on the small-distance behavior.
  Diffusion coefficients are scaled by $\kT/\etam$ and the distance by
  the Saffman-Delbr\"uck length $\kappa^{-1}$. The full behavior
  [numerical integration of Eq.\ (\ref{gij}) using $\kappa h=10^2$,
  solid] is shown together with those in the various asymptotic
  regions: free near region ($\Dlt\sim\ln r$, dashed); free far region
  ($\Dl\sim 1/r$, $\Dt\sim 1/r^2$, dash-dotted); supported region
  ($\Dlt\sim\pm 1/r^2$, dotted).}
\label{fig_Dh}
\end{figure}

The dependencies of the coupling diffusion coefficients on the
separation between the inclusions for the hovering regime, along with
the various asymptotic regions, are shown in Fig.\ \ref{fig_Dh}.

\section{Effective response}
\label{sec_effective}

The presence of inclusions in the membrane influences its response to
stresses. In regular suspensions the response far from the point of
perturbation is similar to that of the particle-free liquid but with a
different prefactor, depending on the volume fraction of particles,
$\phi$. The modified prefactor defines a modified, effective
viscosity, $\eta\rightarrow\etaeff(\phi)$. For a 3D suspension of hard
spheres the effective viscosity, to leading order in $\phi$, was
calculated by Einstein \cite{Einstein} as $\etaeff=\eta[1+(5/2)\phi]$.
The analogous calculation for a 2D suspension of hard disks yields
\cite{Belzons} $\etaeff=\eta\left(1+2\phi\right)$. As we have seen in
Sec.\ \ref{sec_model}, the inclusion-free supported membrane may
exhibit a 2D-like response, a 3D-like one, or neither of the
two. Hence, one expects a more complex modification of the response
due to the presence of inclusions, depending on the various lengths in
the problem.  For sufficiently small distances momentum is conserved
within the membrane, and we expect the response to be 2D-like. For
sufficiently large distances transverse momentum is lost to the
substrate, and the membrane is expected to behave neither as a 3D
suspension nor as a 2D one. Heuristically, and also based on the
results for a free membrane \cite{bpj09}, we anticipate that the
effective response of the supported membrane, having an area fraction
$\phi$ of disklike inclusions, $\tenG\rightarrow\tenG^{\rm eff}$, will
be obtained by the following transformation of the parameters:
\begin{eqnarray}
  \etam &\rightarrow& \etameff(\phi) \simeq \etam (1+2\phi) \nonumber\\
  \kappa &\rightarrow& \kappa^{\rm eff}(\phi) = 
   2\etaf/\etameff \simeq \kappa (1-2\phi) \nonumber\\
  \alpha &\rightarrow& \alpha^{\rm eff}(\phi) = \left[\kappa^{\rm eff}/(2 h)\right]^{1/2} 
  \simeq \alpha (1-\phi). 
\label{transformation}
\end{eqnarray}
We now proceed to prove that this is indeed the case.

Let us begin with an inclusion-free membrane and apply an in-plane,
localized force density, $\vecF\delta(\vecr)$, at the origin. The
resulting flow velocity field of the membrane is given by
$v^{(0)}_i(\vecr)=G_{ij}(\vecr)F_j$, where $\tenG(\vecr)$ is the
Green's function discussed in Sec.\ \ref{sec_model}, Eq.\
(\ref{gij}). Next, let us consider the change in velocity at position
$\vecr$, $\delta\vecv(\vecr,\vecr')$, due to a single disklike
inclusion located at $\vecr'$.  No force or torque are acting on the
inclusion and, hence, its leading correction to the flow velocity is
through the force dipole (stresslet) $\tenS$ that it introduces,
$\delta v_i=S_{kj}(\vecr')\pd_kG_{ij}(\vecr-\vecr')$. There is a local
relation between $\tenS(\vecr')$ and the inclusion-free velocity field
at $\vecr'$, given by
\begin{equation}
  S_{ij} = 2\pi\etam a^2 \left( 1+\frac{1}{8} a^2 \nabla^2\right)
  \left(\partial_j v^{(0)}_i + \partial_i v^{(0)}_j\right).
\label{force_dipole}
\end{equation}
This membrane-analogue of Fax\'en's second relation \cite{Faxen} was
derived in Ref.\ \cite{bpj09} for a free membrane under the assumption
$\kappa a\ll 1$. It remains valid in the current case of a supported
membrane, provided that $a\ll\min(\kappa^{-1},\alpha^{-1})$.

In the next stage we consider randomly distributed inclusions,
occupying an area fraction $\phi$ of the membrane.  We restrict the
calculation to the leading (linear) order in $\phi$, where static
correlations as well as hydrodynamic interactions between inclusions
can be neglected. The average correction to the flow velocity is given
then by integration over all possible positions $\vecr'$ of
inclusions, multiplied by the uniform probability density of finding
an inclusion centered at that position, $\phi/(\pi a^2)$,
\begin{equation}
  \langle \delta v_i(\vecr)\rangle =  \frac{\phi}{\pi a^2} \int d^2 r'
  S_{kj}(\vecr') \pd_k G_{ij}(\vecr-\vecr').
\label{deltav(r)}
\end{equation}
The convolution in Eq.\ (\ref{deltav(r)}) is conveniently handled in Fourier space,
\begin{equation}
  \langle \delta \tilde{v}_i(\vecq)\rangle = \frac{\phi}{\pi a^2} \tilde{S}_{kj}(\vecq) 
  \imath q_k \tG_{ij}(\vecq) = -2\phi \frac{2q/\kappa} {\coth(qh)+2q/\kappa+1} 
  \tG_{ij}(\vecq) F_j,
\label{deltav(q)}
\end{equation}
where in the last equation we have used Eqs.\ (\ref{Gq}) and
(\ref{force_dipole}) while neglecting the term of order $(qa)^2$ in
Eq.\ (\ref{force_dipole}). Writing $\tilde{v}_i=\tilde{v}^{(0)}_i +
\langle\delta\tilde{v}_i\rangle = \tG^{\rm eff}_{ij}F_j$, we identify
the effective Green's function as
\begin{equation}
  \tilde{\tenG}^{\rm eff} = \left[ 1 - 2\phi \frac{2q/\kappa} 
  {\coth(qh)+2q/\kappa+1} \right]
  \tilde{\tenG}.
\label{tGeff}
\end{equation}
It is readily verified that the same result is obtained from Eq.\
(\ref{Gq}) through the transformation defined in Eq.\
(\ref{transformation}) and expansion to linear order in $\phi$.

The $q$-dependence of the renormalization factor in Eq.\ (\ref{tGeff})
reflects the aforementioned complex response of the
inclusion-decorated membrane. In the $q\rightarrow\infty$ limit the
prefactor becomes $1-2\phi$, as in a 2D suspension \cite{Belzons}. In
the opposite limit of $q\rightarrow 0$ the prefactor tends to unity ---
\ie the inclusions have no effect on the large-distance velocity
response of the membrane. To analyze the effective response in more
detail, away from these two limits, one may invert Eq.\ (\ref{tGeff})
back to real space in the desired limit.  Alternatively, one can
substitute the transformation defined in Eq.\ (\ref{transformation})
in the various limiting expressions, already derived for
$\tenG(\vecr)$ in Sec.\ \ref{sec_model} [Eqs.\ 
(\ref{Gra})--(\ref{Gsup})], and expand those expressions to linear
order in $\phi$.  Either of these two procedures yields the effective
response of the membrane in the various regimes listed in Table
\ref{tab_regime}.
We now examine the resulting expressions for the different regimes.

In the adsorbed regime, $h\ll \kappa^{-1}$, the substrate
affects the response of the membrane.  The correction to the Green's
function of Eq.\ (\ref{Gra}) due to the presence of inclusions is
given in this regime by
\begin{eqnarray}
  h\ll\kappa^{-1}:\ \ && \tenG^{\rm eff} \simeq \tenG^{\rm a} + \delta\tenG^{\rm a} 
 \nonumber\\
  && \delta G^{\rm a}_{ij}(\vecr) = \frac{\phi}{2\pi\etam} 
  \left[ \left(-K_0(\alpha r)+\alpha r K_1(\alpha r) \right)\delta_{ij} - 
  \alpha r  K_1(\alpha r)\frac{r_ir_j}{r^2}\right].
\label{Ga_eff}
\end{eqnarray}
In the hovering free regions, $\kappa^{-1}\ll h$ and $r\ll h$,
the membrane is insensitive to the presence of the substrate.  The
correction to the Green's function of Eq.\ (\ref{Grf}) for these free
regions is
\begin{eqnarray}
  \kappa^{-1}\ll h:\ \ && \tenG^{\rm eff} \simeq \tenG^{\rm f} + \delta\tenG^{\rm f} \nonumber\\
  && \delta G^{\rm f}_{ij}(\vecr) = \frac{\phi}{2\etam} \left\{
  \frac{1-(\kappa r)^2}{(\kappa r)^2} \left[ \frac{2}{\pi(\kappa r+1)}
  + \kappa r [ H_{-1}(\kappa r) + Y_1(\kappa r)] \right] \delta_{ij} \right. \nonumber \\
  && +\left. \left[ \frac{4(\kappa r-1)}{\pi(\kappa r)^2} + 
  \frac{(\kappa r)^2 - 2}{\kappa r} [ H_{-1}(\kappa r) + Y_1(\kappa r)]
  - H_0(\kappa r)  + Y_0(\kappa r) \right] \frac{r_ir_j}{r^2} \right\}.
\label{Gf_eff}
\end{eqnarray}

These two limiting cases (adsorbed and hovering free) are further
subdivided into near and far regions.  The near regions of both cases
are governed by 2D shear stresses, leading to an effective response
similar to that of a 2D particulate liquid with a leading logarithmic
behavior.  The differences arise from the different cutoff lengths ---
$\alpha^{-1}$ in the adsorbed near region and $\kappa^{-1}$ in the
free near region.  The different dependencies of these lengths on the
membrane viscosity lead to slightly different concentration
corrections as $\etam$ is modified to $\etameff$. In the adsorbed near
region, $r\ll\alpha^{-1}\ll\kappa^{-1}$, the correction to the Green's
function of Eq.\ (\ref{Gan}) is given by
\begin{eqnarray}
  &&h\ll\kappa^{-1}, r\ll\alpha^{-1}: \nonumber\\ 
  &&\tenG^{\rm eff} \simeq (1-2\phi)G^{\rm an}_{ij} + \frac{\phi}{4\pi\etam}\delta_{ij} \nonumber\\
  &&\ \ \ \ \ \ 
  = \frac{1}{4\pi\etam} \left\{ \left[-(1-2\phi)
    \left( \ln(\alpha r/2) + \gamma + 1/2 \right) + \phi \right] \delta_{ij}
  + (1-2\phi) \frac{r_ir_j}{r^2} \right\}.
\label{Gan_eff}
\end{eqnarray}
For the free near region, $r\ll\kappa^{-1}\ll h$, the Green's function
of Eq.\ (\ref{Gfn}) is modified according to
\begin{eqnarray}
  &&r\ll\kappa^{-1}\ll h: \nonumber\\
  &&\tenG^{\rm eff} \simeq (1-2\phi)G^{\rm
  fn}_{ij} + \frac{\phi}{2\pi\etam}\delta_{ij} \nonumber\\ 
  &&\ \ \ \ \ \ 
  = \frac{1}{4\pi\etam} \left\{ \left[-(1-2\phi) \left( \ln(\kappa r/2)
  + \gamma + 1/2 \right) + 2\phi \right] \delta_{ij} + (1-2\phi)
  \frac{r_ir_j}{r^2} \right\}.
\label{Gfn_eff}
\end{eqnarray}
Equation (\ref{Gfn_eff}) has already been derived in Ref.\ \cite{bpj09}
for a free membrane.

For all other regions we find no modification of the dominant term in
the membrane response due to the presence of inclusions. This is
because the response in these regions is governed by mechanisms which
are unrelated to the propagation of 2D shear stresses in the membrane
and, therefore, insensitive to membrane properties. In the adsorbed
far region, $h\ll\kappa^{-1}$ and $r\gg\alpha^{-1}$, and the supported
region, $\kappa^{-1}\ll h\ll r$, shear stresses are lost to the
substrate. The remaining compressive effect (effective mass dipole) is
insensitive to the presence of inclusions, since the membrane is
assumed to be incompressible. Thus, the dominant response in these
regions is given by the unperturbed Eqs.\ (\ref{Gaf}) and
(\ref{Gsup}).  In the free far region, $\kappa^{-1}\ll r\ll h$, the
response is dominated by 3D shear stresses in the adjacent fluid,
which are obviously indifferent to the inclusions.  Hence, in this
region the dominant response remains equal to the unperturbed Eq.\
(\ref{Gff}).

In the cases where the leading membrane response is not renormalized
by the inclusions, there are nevertheless higher-order corrections
which do depend on $\phi$.  To conclude this section we address these
large-distance corrections. In the adsorbed far region the correction
to Eq.\ (\ref{Gaf}) is exponentially small in $\alpha r$ and,
therefore, negligible. In the free far region the leading correction
to Eq.\ (\ref{Gff}) is obtained from Eq.\ (\ref{Gf_eff}) in the limit
$\kappa r\gg 1$ as
\begin{eqnarray}
  \kappa^{-1}\ll r\ll h:\ \ && \tenG^{\rm eff} \simeq \tenG^{\rm ff} + \delta\tenG^{\rm ff} 
 \nonumber\\
  && \delta G^{\rm ff}_{ij}(\vecr) = \frac{\phi}{\pi\etam (\kappa r)^2}
  \left( \delta_{ij} - \frac{2r_ir_j}{r^2} \right).
\label{Gff_eff}
\end{eqnarray}
Thus, the correction is of higher order [$1/r^2$ compared to $1/r$ in
Eq.\ (\ref{Gff})] but still long-ranged. In the supported region we
substitute in Eq.\ (\ref{tGeff}) the expansion $\coth(qh)+2q/\kappa\simeq
(qh)^{-1}+3qh$ and invert back to real space while assuming the limit
$r\gg h$. This procedure yields the following correction to Eq.\ 
(\ref{Gsup}):
\begin{eqnarray}
  r\gg h\gg\kappa^{-1}:\ \ && \tenG^{\rm eff} \simeq \tenG^{\rm s} + \delta\tenG^{\rm s} 
 \nonumber\\
  && \delta G^{\rm s}_{ij}(\vecr) = \frac{12\phi h^3}{\pi\etaf\kappa r^5}
  \left( 4\delta_{ij} - \frac{5r_ir_j}{r^2} \right),
\label{Gsup_eff}
\end{eqnarray}
which decays algebraically but much faster than the unperturbed
response ($1/r^5$ vs.\ $1/r^2$).

\section{Corrected pair-diffusion coefficients}
\label{sec_corrected}      

Substituting in Eq.\ (\ref{DltG}) the effective response functions,
calculated in Sec.\ \ref{sec_effective}, readily gives the corrections
to the coupling diffusion coefficients to leading order in the area
fraction of inclusions,
\begin{equation}
  D_{\rm L,T}^{\rm eff} \simeq \Dlt + \delta\Dlt,
\end{equation}
where $\Dlt$ are the inclusion-free coefficients derived in Sec.\ 
\ref{sec_diffusion}. We now provide the resulting expressions for $\delta\Dlt$ in
the various spatial regimes.

\subsection{Adsorbed regime: $h\ll\kappa^{-1}$} 
\label{sec_corrected_a}

In the adsorbed regime we use $\delta\tenG^{\rm a}$ of Eq.\ (\ref{Ga_eff}) in 
Eq.\ (\ref{DltG}) to get
\begin{eqnarray}
  h\ll\kappa^{-1}:\ \ &&\delta\Dl(r) \simeq -\phi \frac{\kT}{2\pi\etam} K_0(\alpha r)
 \nonumber\\
  &&\delta\Dt(r) \simeq -\phi \frac{\kT}{2\pi\etam} 
  \left[K_0(\alpha r)-\alpha rK_1(\alpha r)   \right].
\label{Da_eff}
\end{eqnarray}
These expressions are the corrections to the bare coefficients given
in Eq.\ (\ref{Dlta}). In the adsorbed near region they reduce to
\begin{equation}
  h\ll\kappa^{-1}:\ \ \delta\Dlt(r\ll\alpha^{-1}) \simeq 
 \phi \frac{\kT}{2\pi\etam} \left[ \ln(\alpha r/2) + \gamma + 1/2 \mp 1/2 \right],
\label{Dan_eff}
\end{equation}
where the upper (lower) sign corresponds to the longitudinal
(transverse) coefficient. Equation (\ref{Dan_eff}) gives the
correction to the bare coefficients of Eq.\ (\ref{Dlt_an}). In the
adsorbed far region, $r\gg\alpha^{-1}$, the corrections are
exponentially small. More specifically, $\delta\Dl\sim (\alpha
r)^{-1/2}e^{-\alpha r}$ and $\delta\Dt\sim (\alpha r)^{1/2}e^{-\alpha
r}$. 

\begin{figure}[tbh]
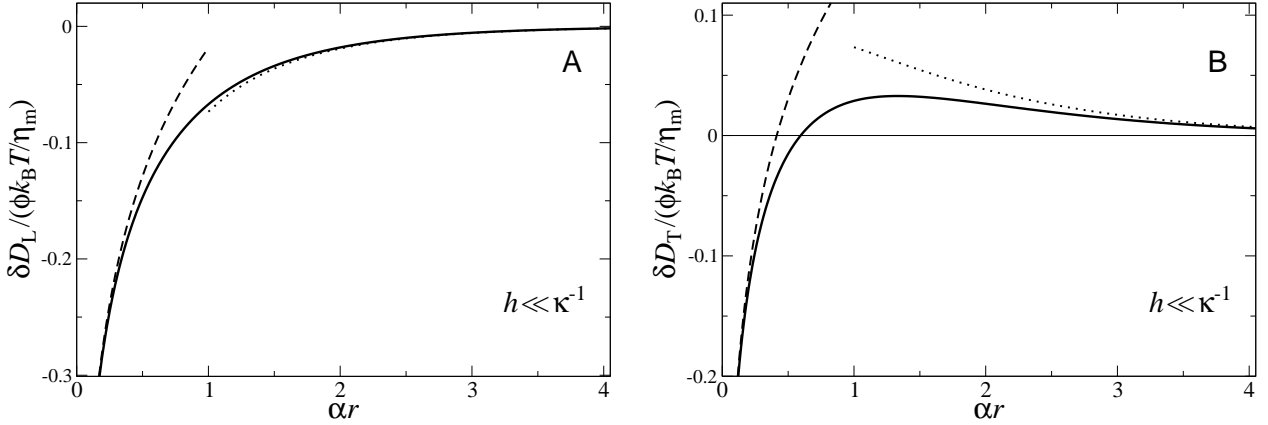

\vspace{0.7cm} 
\centerline{\resizebox{0.45\textwidth}{!}{\includegraphics{fig4a.eps}}
\hspace{0.3cm}
\resizebox{0.45\textwidth}{!}{\includegraphics{fig4b.eps}}}
\caption{Corrections to the longitudinal (A) and transverse (B)
  coupling diffusion coefficients as a function of interparticle
  distance for the adsorbed regime ($h\ll\kappa^{-1}$).  The
  corrections are scaled by $\phi\kT/\etam$ and the distance by the
  momentum-screening length $\alpha^{-1}$. The full behavior [Eq.\ 
  (\ref{Da_eff}), solid] is shown together with those in the two
  asymptotic regions: adsorbed near region ($\delta\Dlt\sim\ln r$,
  dashed); adsorbed far region ($\delta\Dlt\sim\mp(\alpha r)^{\mp
    1/2}e^{-\alpha r}$, dotted).}
\label{fig_deltaDa}
\end{figure}

The behavior of the concentration corrections as a function of
separation in the adsorbed regime is shown in Fig.\ \ref{fig_deltaDa}.
Notice the range of positive correction to $\Dt$, \ie the unusual {\it
  increase} in the transverse coupling due to the presence of
inclusions.

\subsection{Hovering regime: $\kappa^{-1}\ll h$} 
\label{sec_corrected_h}

In the hovering free regions, $\kappa^{-1}\ll h$ and $r\ll h$, we
substitute $\delta\tenG^{\rm f}$ of Eq.\ (\ref{Gf_eff}) in Eq.\ 
(\ref{DltG}) to obtain
\begin{eqnarray}
  &&\kappa^{-1}\ll h:\nonumber\\ 
  &&\delta\Dl(r\ll h) \simeq -\phi\frac{\kT}{2\etam}
   \left[ \frac{2}{\pi(\kappa r)^2} + H_0(\kappa r)
   - \frac{H_1(\kappa r)}{\kappa r} + \frac{1}{2} \left(
   Y_2(\kappa r) - Y_0(\kappa r) \right) \right]
 \nonumber\\
  && \delta\Dt(r\ll h) \simeq \phi\frac{\kT}{2\etam} 
   \frac{(1-(\kappa r)^2)}{\kappa r} \left[
   \frac{2}{\pi\kappa r (1+\kappa r)} + H_{-1}(\kappa r) + Y_1(\kappa r)
   \right],
\label{Df_eff}
\end{eqnarray}
which are the corrections to the coefficients of Eq.\
(\ref{Dlt_f}). The hovering free behavior is further subdivided into
near and far regions. In the free near region, the expressions in Eq.\
(\ref{Df_eff}) become
\begin{equation}
  \kappa^{-1}\ll h:\ \ \delta\Dlt(r\ll\kappa^{-1}) \simeq 
  \phi \frac{\kT}{2\pi\etam} \left[ \ln(\kappa r/2) + \gamma+ 1 \mp 1/2 \right],
\label{Dfn_eff}
\end{equation}
which corrects Eq.\ (\ref{Dlt_fn}). In the free far region Eq.\
(\ref{Df_eff}) reduces to
\begin{equation}
   \kappa^{-1}\ll h:\ \ \delta\Dlt(\kappa^{-1}\ll r\ll h) 
  \simeq \mp \phi \frac{\kT}{\pi\etam} \frac{1}{\kappa^2r^2},
\label{Dff_eff}
\end{equation}
which are the corrections to $\Dl$ and $\Dt$ of Eq.\
(\ref{Dlt_ff}). Note that in the free far region $\Dt$ and $\delta\Dt$
both decay as $1/r^2$, whereas $\Dl$ has a slower decay ($\sim 1/r$)
than its correction ($\sim 1/r^2$). Thus, the longitudinal coefficient
in the free far region remains essentially unaffected by the
inclusions. The results for the hovering free region coincide with
those derived in Ref.\ \cite{bpj09} for a free membrane.

In the last region, the supported region, where $r\gg
h\gg\kappa^{-1}$, the dominant term in the large-distance response
[Eq.\ (\ref{Gsup})] is insensitive to the properties of the membrane.
The corrections to the coupling diffusion coefficients of Eq.\ 
(\ref{Dlt_hs}), therefore, are of higher order. Substituting
$\delta\tenG^{\rm s}$ of Eq.\ (\ref{Gsup_eff}) in Eq.\ (\ref{DltG}),
we get
\begin{equation}
  \kappa^{-1}\ll h:\ \ \ 
  \delta\Dl(r\gg h) \simeq -\phi \frac{12\kT h^3}{\pi\etaf\kappa r^5},\ \ \  
  \delta\Dt(r\gg h) \simeq  \phi\frac{48\kT h^3}{\pi\etaf\kappa r^5}.
\end{equation}

\begin{figure}[tbh]
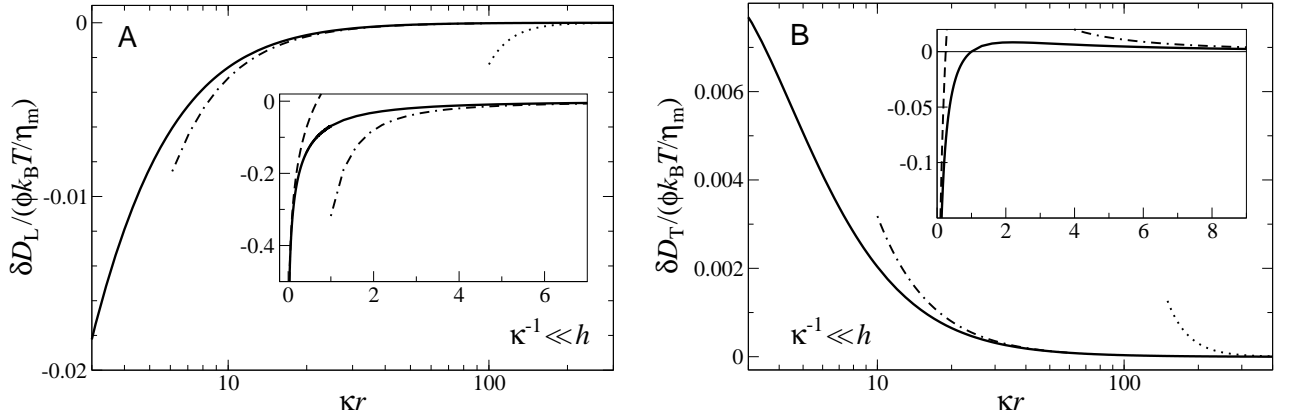

\vspace{0.7cm} 
\centerline{\resizebox{0.45\textwidth}{!}{\includegraphics{fig5a.eps}}
\hspace{0.3cm}
\resizebox{0.46\textwidth}{!}{\includegraphics{fig5b.eps}}}
\caption{Corrections to the longitudinal (A) and transverse (B) 
  coupling diffusion coefficients as a function of interparticle
  distance for the hovering regime ($\kappa^{-1}\ll h$). Insets focus
  on the small-distance behavior. Corrections are scaled by
  $\phi\kT/\etam$ and the distance by the Saffman-Delbr\"uck length
  $\kappa^{-1}$. The full behavior [numerical inversion of Eq.\ 
  (\ref{tGeff}) using $\kappa h=10^2$, solid] is shown together with
  the asymptotes for the free near region ($\delta\Dlt\sim\ln r$,
  dashed) and free far region ($\delta\Dlt\sim \mp 1/r^2$,
  dash-dotted). At much larger distances the supported region
  ($\delta\Dlt\sim \mp 1/r^5$, dotted) sets in.}
\label{fig_deltaDh}
\end{figure}

The spatial behavior of the concentration corrections as a function of
separation in the hovering regime is shown in Fig.\
\ref{fig_deltaDh}. Notice again the broad range of positive correction
to $\Dt$, where the transverse coupling {\it increases} with the
concentration of inclusions. This is a consequence of the unusual
dependence of the bare coefficient on membrane viscosity. [See Eq.\
(\ref{Dlt_ff}) and the text below it.]

\section{Correlated diffusion of large inclusions}
\label{sec_large}

The entire analysis so far has relied on the assumption that the
inclusion size is much smaller than any other length in the system,
$a\ll\min(\kappa^{-1},h,r)$. Consequently, the coupling diffusion
coefficients derived in Sec.\ \ref{sec_diffusion} were independent of
the size and shape of the inclusions. In the adsorbed far region
($r\gg\alpha^{-1}\gg h$) we can depart from this assumption and derive
the large-separation coupling diffusion coefficients for large
disklike inclusions.  In principle, the calculation of pair mobilities
for two large particles is technically hard --- one needs to solve the
flow equations with boundary conditions on the surfaces of the two
particles as they move with prescribed velocities (or under prescribed
forces). For the adsorbed far region, however, there is a scheme that
bypasses this difficulty altogether. It is based on symmetry
considerations and the knowledge of the exact flow field away from a
single moving disk.

In the adsorbed regime our model becomes equivalent to an effective 2D
Brinkman fluid \cite{Brinkman} --- \ie an incompressible fluid with
momentum decay --- as previously studied in Refs.\ 
\cite{Mazenko}--\cite{Komura}. This is clearly seen in the
corresponding velocity response function, Eq.\ (\ref{Gqa}), which
contains a momentum decay length, $\alpha^{-1}$.

First, let us recall, for the adsorbed regime, the far flow in the
membrane due to a point force $\vecF$.  According to Eq.\ (\ref{Gaf})
this flow is the same as the one emanating from an effective 2D mass
dipole of strength $(h/\etaf)\vecF$.  Now consider the flow due to a
force $\vecF$ applied to an isolated disk of radius $a_1$, positioned
at the origin.  This problem was solved for a 2D Brinkman fluid in
Ref.\ \cite{ES}. Applying the result to our case, we get in the far
field [$r\gg\max(a_1,\alpha^{-1})$] the following dipolar flow:
\begin{eqnarray}
  v_i(\vecr) &=& -\frac{h}{\etaf} \frac{m(\alpha a_1)} {2\pi r^2} 
   \left(\delta_{ij} - \frac{2r_ir_j}{r^2} \right)F_j \nonumber\\
  m(x) &=& 2 \frac{xK_0(x) + 2K_1(x)} {xK_0(x) + 4K_1(x)}.
\label{diskdipole}
\end{eqnarray}
Thus, no matter how large the disk may be, the far flow remains
equivalent to that induced by a 2D mass dipole; the only dependence on
the particle size is through the dimensionless prefactor $m(\alpha
a)$.  

In the next step we identify the tensor multiplying $\vecF$ in Eq.\
(\ref{diskdipole}) with the coupling mobility of two different
particles --- one of radius $a_1$ at the origin and another of
vanishing radius, $a_2\rightarrow 0$, at $\vecr$.  Due to the symmetry
of the coupling mobility, the same tensor also gives the velocity of a
disk of radius $a_1$, positioned at $\vecr$, due to a point force
$\vecF$ applied to the membrane at the origin. Yet, the latter is the
velocity acquired by a particle of radius $a_1$ as it is embedded in a
flow caused by a mass dipole of strength $(h/\etaf)\vecF$. All we
need to do now to get the velocity of a disk of radius $a_1$ due to a
force $\vecF$ exerted on a sufficiently distant disk of radius $a_2$
is to increase the mass-dipole strength at the origin from
$(h/\etaf)\vecF$ up to $(h/\etaf)m(\alpha a_2)\vecF$, the
effective mass dipole created by $\vecF$ when it is applied to a
particle of radius $a_2$.  Hence, the large-distance coupling mobility
is given by
\begin{equation}
  B_{12,ij}(\vecr) = -\frac{h}{\etaf} \frac{m(\alpha a_1)m(\alpha a_2)}
   {2\pi r^2} \left(\delta_{ij} - \frac{2r_ir_j}{r^2} \right).
\label{Blarge}
\end{equation}

As explained in Sec.\ \ref{sec_diffusion}, the coupling diffusion
coefficients can be readily obtained from Eq.\ (\ref{Blarge}) as
$\Dl(r)=\kT B_{12,xx}(r\xhat)$ and $\Dt(r)=\kT B_{12,yy}(r\xhat)$,
yielding
\begin{equation}
  r\gg\max(\alpha^{-1},a_1,a_2):\ \ 
  \Dlt(r) \simeq \pm \frac{\kT m(\alpha a_1)m(\alpha a_2) h} {2\pi\etaf r^2},
\label{Dltlarge}
\end{equation}
where the plus (minus) sign corresponds to the longitudinal
(transverse) coefficient, and $m(x)$ is defined in Eq.\ 
(\ref{diskdipole}).  Equation (\ref{Dltlarge}) gives the
large-distance coupling diffusion coefficients in the adsorbed regime
for two disklike inclusions of arbitrary radii.

In the limit of small inclusions, $\alpha a_\beta\ll 1$ ($\beta=1,2$),
we have $m(\alpha a_\beta)\simeq 1$, and the result of Sec.\
\ref{sec_diffusion} [Eq.\ (\ref{Dlt_hs})] is recovered. In the
opposite limit of large inclusions, $m(\alpha a_\beta)\simeq 2$, and,
therefore,
\begin{equation}
  r\gg a_\beta\gg \alpha^{-1}:\ \ 
   \Dlt(r) \simeq \pm \frac{2\kT h} {\pi\etaf r^2}.
\end{equation}
Thus, going from small to large inclusions changes the large-distance
coupling coefficients by a mere factor of $4$. Interestingly, the
results for very large inclusions are again independent of particle
size and shape. We return to this surprising finding in the next
section.

\section{Conclusions}
\label{sec_conclusion}

The aim of this work has been to characterize supported membranes as
effective heterogeneous fluids. The existence of several length scales
in the problem leads to various regimes that are governed by different
physical mechanisms and exhibit different effective dimensionality
(either 2D or 3D); see Table \ref{tab_regime}. We have provided
predictions for the coupling diffusion coefficients of inclusion pairs
in those various regimes, as well as their leading dependence on the
concentration of inclusions in the membrane. These predictions can be
directly checked in two-point microrheology experiments using Eq.\ 
(\ref{fluct}).

Since the SD length, $\kappa^{-1}$, is typically of micron scale,
common supported membranes should belong in the adsorbed regime, $h\ll
\kappa^{-1}$, which is treated in Secs.\ \ref{sec_diffusion_a} and
\ref{sec_corrected_a}. Moreover, the limit of small inclusion size,
assumed in those sections, should be generally valid, since the
requirement is that $a$ be much smaller than
$\alpha^{-1}\sim(\kappa^{-1}h)^{1/2}$ rather than the stricter condition
$a\ll h$. Hence, we expect this limit to hold for common membrane
inclusions even in cases where the distance to the substrate is of the
order of the inclusion size (say, a few nm only). In this common
scenario of $a\ll\alpha^{-1}\ll\kappa^{-1}$ the substrate is predicted
to strongly suppress the large-distance correlations as compared to a
free membrane.  Comparing the results of Sec.\ \ref{sec_diffusion_a}
with those of Ref.\ \cite{bpj09} (or with the equivalent results for
the free far region in Sec.\ \ref{sec_diffusion_h}), we find
suppression of the longitudinal and transverse coefficients by factors
of order $h/r$ and $\kappa h$, respectively. Nonetheless, the
correlations always remain long-ranged, their fastest possible decay
being as $1/r^2$.

For nm-scale separation between membrane and substrate, which is
comparable to the membrane thickness, the substrate may affect the
membrane properties. As long as the membrane remains fluid, such
interactions are expected to merely modify the effective membrane
viscosity, and the theory presented here should remain valid. A more
serious concern is the possible breakdown of the bilayer description
as a uniform slab, which is inherent in the Saffman-Delbr\"uck model
and the current work. At sufficient proximity to the surface the
dynamics of the two membrane leaflets might decouple. This will occur
when the friction between the lower leaflet and the solid surface
exceeds the one between the two leaflets. The characteristic
coefficients for these two competing drags are, respectively,
$\etaf/h\sim 10^6$ N$\cdot$s/m$^3$ (for $h\sim 1$ nm) and $10^8$
N$\cdot$s/m$^3$ \cite{YeungEvans}. Thus, for all relevant separations
$h$, relative motion of the leaflets should not play a significant
role, and the bilayer can be considered as a single fluid medium.

There may be cases where the inclusion size is comparable to or larger
than $\alpha^{-1}$ --- for example, when the inclusion is a colloid
particle or a membrane domain. We have presented expressions for
the large-distance coupling coefficients in this case as well (Sec.\ 
\ref{sec_large}). These results have been derived for the specific
case of disklike inclusions, yet in both limits of small and large
$\alpha a$ they become independent of the size and shape of the
inclusions. The origin of this surprising universality is that, in the
adsorbed regime, the membrane responds to any size and shape of
perturbation sufficiently far away, as if the perturbation were a
mass dipole. The `effective inclusion' --- \ie the region around the
perturbation whose dynamics determines the strength of that mass
dipole --- is limited in both cases of very small and very large $a$ by
the momentum decay length $\alpha^{-1}$.

Concerning the effective response of the supported membrane as a
function of the area fraction of inclusions, we have found that the
membrane viscosity is modified according to the law for 2D suspensions
\cite{Belzons}, $\etameff=\etam(1+2\phi)$, yet this modification should
be included also in the parameters $\kappa$ and $\alpha$ [Eq.\ 
(\ref{transformation})]. The combined effect is that there is {\em no}
renormalization of the large-distance membrane response with
increasing $\phi$. The underlying physics is that transverse momentum
is not transferred through the membrane over large distances and,
hence, the response is insensitive to changes in the membrane
viscosity.  This insensitivity holds already for distances much
smaller than the largest length scale in the problem at hand. For the
common, adsorbed regime, $h\ll\kappa^{-1}$, it is valid for
$r\gg\alpha^{-1}$, as momentum is first lost to the substrate. For the
hovering regime, $\kappa^{-1}\ll h$, the dependence on $\phi$
disappears for $r\gg\kappa^{-1}$, as the propagation of stresses becomes
dominated by the outer fluid.

The subtle effect of increasing the area fraction of inclusions on the
membrane viscosity and the governing length scales influences also the
corrections to the coupling diffusion coefficients (Sec.\ 
\ref{sec_corrected}). At sufficiently short distances the leading
logarithmic terms are corrected as if the membrane were a 2D
suspension, yet at large distances all corrections to the coupling
coefficients vanish.

The validity of the theory presented here is limited in several
important respects. Our results concerning the coupling diffusion
coefficients are all valid only for large separations, $r\gg a$. In
addition, the coupling coefficients derived in Sec.\ \ref{sec_large}
for large inclusions apply only in the adsorbed far region,
$r\gg\alpha^{-1}\gg h$; the corresponding expressions for large
inclusions in the other asymptotic regions are unknown. We have
restricted the analysis of the effective response and corrections to
the coupling coefficients to the leading linear order in the area
fraction of inclusions, $\phi$.  Deviations from the theory are
expected, therefore, as $\phi$ becomes appreciable. Nevertheless, as
is clear from the discussion above, our main qualitative results ---
in particular, the insensitivity of the large-distance response to
$\phi$ --- are expected to be valid for all values of $\phi$, so long
as the membrane remains fluid.  Finally, we have not considered
membrane fluctuations, which may have a subtle interplay with the
diffusion of membrane inclusions \cite{Brown,Seifert}. As membrane
fluctuations are suppressed by the presence of a nearby surface
\cite{SunilKumar,GovZilman}, we do not expect our main results to be
significantly influenced by such effects.

\begin{acknowledgments}
  We are indebted to Shigeyuki Komura and Sanoop Ramachandran for
  pointing out an error in the original manuscript and for sharing
  their results prior to publication. We thank Matan Ben-Zion, Frank
  Brown, Gilad Haran, and Maria Ott for helpful discussions.  This
  research has been supported by the Israel Science Foundation (Grant
  No.\ 588/06).
\end{acknowledgments}


\end{document}